\documentclass[12pt,a4paper]{article}
\pdfoutput=1

\usepackage[utf8]{inputenc}
\usepackage[T1]{fontenc}

\usepackage{multirow}
\usepackage{amsmath}
\usepackage{color}
\usepackage{amsfonts}
\usepackage{amssymb}
\usepackage{graphicx}
\usepackage{geometry}
\usepackage{amssymb,epsfig,subfigure}
\usepackage{hyperref}
\usepackage{url}
\usepackage{comment}
\usepackage[font=footnotesize]{caption}
\usepackage{slashed}

\makeatletter
\renewcommand\section{\@startsection {section}{1}{\z@}%
                                 {-3.5ex \@plus -1ex \@minus -.2ex}
                                   {2.3ex \@plus.2ex}%
                                   {\normalfont\large\bfseries}}
\renewcommand\subsection{\@startsection{subsection}{2}{\z@}%
                                   {-3.25ex\@plus -1ex \@minus -.2ex}%
                                     {1.5ex \@plus .2ex}%
                                     {\normalfont\bfseries}}
\renewcommand\subsubsection{\@startsection{subsubsection}{3}{\z@}%
                                   {-3.25ex\@plus -1ex \@minus -.2ex}%
                                     {1.5ex \@plus .2ex}%
                                     {\normalfont\itshape}}
\makeatother

\def\pplogo{\vbox{\kern-\headheight\kern -29pt
\halign{##&##\hfil\cr&{\ppnumber}\cr\rule{0pt}{2.5ex}&\ppdate\cr}}}
\makeatletter
\def\ps@firstpage{\ps@empty \def\@oddhead{\hss\pplogo}%
  \let\@evenhead\@oddhead 
}
\def\maketitle{\par
 \begingroup
 \def\thefootnote{\fnsymbol{footnote}}
 \def\@makefnmark{\hbox{$^{\@thefnmark}$\hss}}
 \if@twocolumn
 \twocolumn[\@maketitle]
 \else \newpage
 \global\@topnum\z@ \@maketitle \fi\thispagestyle{firstpage}\@thanks
 \endgroup
 \setcounter{footnote}{0}
 \let\maketitle\relax
 \let\@maketitle\relax
 \gdef\@thanks{}\gdef\@author{}\gdef\@title{}\let\thanks\relax}
\makeatother

\numberwithin{equation}{section}

\newcommand\eea{\end{eqnarray}}
\newcommand\bea{\begin{eqnarray}}

\def\beq{\begin{equation}}
\def\eeq{\end{equation}}

\newcommand{\be}{\begin{equation}}
\newcommand{\ee}{\end{equation}}
\newcommand{\ba}{\begin{align}}
\newcommand{\ea}{\end{align}}
\newcommand{\bg}{\begin{gather}}
\newcommand{\eg}{\end{gather}}
\newcommand{\bseq}{\begin{subequations}}
\newcommand{\eseq}{\end{subequations}}

\newcommand{\coment}[1]{}

\begin{document}
\setcounter{page}0
\def\ppnumber{\vbox{\baselineskip14pt
}}
\def\ppdate{
} \date{}

\author{Horacio Casini$^{1,2}$, Ignacio Salazar Landea$^3$, Gonzalo Torroba$^{1, 2}$\\
[7mm] \\
{\normalsize \it $^1$ Instituto Balseiro, UNCuyo and CNEA}\\
{\normalsize \it $^2$Centro At\'omico Bariloche and CONICET}\\
{\normalsize \it S.C. de Bariloche, R\'io Negro, R8402AGP, Argentina}\\
{\normalsize \it $^3$Instituto de F\'\i sica de La Plata - CONICET, C.C. 67, 1900 La Plata, Argentina}\\
}

\bigskip
\title{\bf  Irreversibility, QNEC, and defects
\vskip 0.5cm}
\maketitle


\begin{abstract}
We first present an analysis of infinitesimal null deformations for the entanglement entropy, which leads to a major simplification of the proof of the  $C$, $F$ and $A$-theorems in quantum field theory. Next, we study the quantum null energy condition (QNEC) on the light-cone for a CFT. Finally, we combine these tools in order to establish the irreversibility of renormalization group flows on planar $d$-dimensional defects, embedded in $D$-dimensional conformal field theories. This proof completes and unifies all known defect irreversibility theorems for defect dimensions $d\le 4$. The F-theorem on defects ($d=3$) is a new result using information-theoretic methods. For $d \ge 4$ we also establish the monotonicity of the relative entropy coefficient proportional to $R^{d-4}$. The geometric construction connects the proof of irreversibility with and without defects through the QNEC inequality in the bulk, and makes contact with the proof of strong subadditivity of holographic entropy taking into account quantum corrections. 
\end{abstract}
\bigskip

\newpage

\tableofcontents

\vskip 1cm

\section{Introduction and summary}\label{sec:intro}

Irreversibility of the renormalization group (RG) expresses the idea that the ultraviolet (UV) and infrared (IR) fixed points of a consistent unitary quantum field theory cannot be arbitrarily selected from the set of conformal invariant models. There are some dimensionless ``renormalization group charges'', that partially characterize the fixed points, and which always decrease along RG flows. 
Irreversibility theorems for unitary relativistic quantum field theories (QFT) have been established for space-time dimensions $d=2,3,4$, the so called $C, F, A$ theorems. At present there are no proofs for higher dimensions.  
     
Another set of related ideas started to develop from the proposal of a ``g-theorem'' \cite{affleck1991universal}, later proved in \cite{friedan2004boundary}. This asserts the decreasing of a defect entropy along the RG for $d=2$ conformal field theories (CFT) coupled with a one dimensional defect. Here there is no bulk RG flow but the defect Hamiltonian provides a physical scale. The CFT in interaction with the defect flows between two different defect fixed points at short and long scales. 
 
 The framework for a natural generalization of this idea is to consider a $D$-dimensional CFT coupled to a $d\le D$ dimensional planar static defect. Relevant interactions on the defect are responsible for triggering an RG flow between UV and IR fixed points, while the bulk remains conformal.      
 Irreversibility in this setup gives generalizations of the $g$-theorem. From the point of view of this paper it will be convenient to think of the irreversibility theorems for QFT without defect as the special case $d=D$ of this more general setup.\footnote{On the contrary, $d=0$ is trivial from the RG point of view since the $D$ dimensional theory is a CFT.} Indeed, for $D=d$ there is no ambient CFT and the theory is just a QFT where the ordinary irreversibility theorems apply. 
 
Though a simple physical understanding on the  necessity of irreversibility is still lacking, the identification of the RG charges at fixed points has been gradually consolidating. For the case $D=d$ (QFT without defect), and for even $d$, it was early pointed out that it should be identified with the Euler term in the conformal anomaly \cite{Zamolodchikov:1986gt,Cardy:1988cwa}, or, equivalently, the logarithmic coefficient of the free energy of the CFT in a Euclidean $d$-dimensional sphere \cite{Cardy:1988cwa}. This turns out to be also equivalent to the logarithmic universal coefficient of the entanglement entropy (EE) of a sphere \cite{holzhey1994geometric,solodukhin2008entanglement,casini2011towards}. 
 For odd $d$, and still $D=d$, where there is no conformal anomaly, it was first identified as the constant term of the entanglement entropy (EE) of a sphere \cite{Myers:2010xs} or the constant term of the free energy of the CFT put in a Euclidean $d$-dimensional sphere \cite{Jafferis:2011zi}. Again, these two descriptions give in fact the same quantity \cite{casini2011towards}.  

For $d=1$, and $D=2$ (the original $g$-theorem), it was identified as the entropy change due to the defect in a thermal state \cite{affleck1991universal}. Again this quantity can be obtained from the entanglement entropy induced by the defect \cite{calabrese2004entanglement} or the constant term in the sphere free energy. A natural generalization for other $D,d$ should then be sought in the universal piece of the EE of a $D-2$ dimensional sphere with a centered defect, or the $D$-dimensional sphere partition function. However, for codimension $D-d\ge 2$ these two quantities do not agree any more. They differ by a energy term in the free energy. The analysis of \cite{Jensen:2018rxu, kobayashi2019towards} indicated that the RG charge is in fact the later one rather than the entropy. In summary, the RG charge always coincides with the universal coefficient of the free energy induced by the defect in a $D$-sphere, that is, the induced constant term for odd $d$ and the induced logarithmic coefficient for even $d$.   

\subsection{Overview of irreversibility inequalities}

Several proofs of irreversibility have being discovered for different $D,d$, and it will be useful to review them briefly. Some of them use traditional QFT methods such as positivity properties of correlation functions and path integrals, or unitarity of the scattering matrix.  Others use quantum information properties of entropies or relative entropies. 

First, non-entropic proofs are listed in Table 1. The $D=d=2$ proof is the original Zamolodchikov's theorem, using reflection positivity (RP) of stress tensor correlators \cite{Zamolodchikov:1986gt}. Starting with \cite{komargodski2011renormalization} for $D=d=4$, all the cases of $d=2$ \cite{jensen2016constraint, Shachar:2022fqk} and $d=4$ \cite{wang2022defect} and any codimension have been proved using dilaton effective actions and either reflection positivity or unitarity of the dilaton scattering matrix. The proof of the case $d=1$ and any codimension \cite{cuomo2022renormalization}, extending the original proof for $d=1,D=2$ in \cite{friedan2004boundary}, uses reflection positivity in the Euclidean path integral.   

\begin{table}[h] 
\begin{center}\begin{tabular}[c]{l| p{2cm}|p{2cm}|p{2cm}|p{2cm}|p{2cm}}
      $ _d\diagdown^D $  & $2$ & $3$ & $4$ & $5$ & $\cdots $\\
      \hline
      1 & \scriptsize{reflection positivity for stress tensor} &   \scriptsize{reflection positivity for stress tensor}  & \scriptsize{reflection positivity for stress tensor}  &  \scriptsize{reflection positivity for stress tensor}  &  \scriptsize{reflection positivity for stress tensor}  \\
      \cline{2-6}
      2 & \scriptsize{reflection positivity for stress tensor} &\scriptsize{reflection positivity of dilaton}& \scriptsize{reflection positivity of dilaton}& \scriptsize{reflection positivity of dilaton} & \scriptsize{reflection positivity of dilaton} \\
       \cline{2-6}
      3 &   & \scriptsize{No proof} & \scriptsize{No proof} &\scriptsize{No proof} & \scriptsize{No proof} \\
      \cline{3-6}
      4 &  \multicolumn{2}{c|}{}  & \scriptsize{unitarity dilaton scattering} & \scriptsize{unitarity  dilaton scattering} & \scriptsize{unitarity  dilaton scattering}\\
       \cline{4-6}
\end{tabular}
\end{center}
\caption{Non entropic proofs according to the origin of the inequality for different $D \ge d$. See text for references.}
\end{table}

The entropic results are summarized in table 2. Entropic proofs exist for $D=d$ and $d=2$ \cite{Casini:2004bw}, $d=3$ \cite{Casini:2012ei}, and $d=4$ \cite{casini2017markov}. All of these use strong subadditivity (SSA) of EE. For $d=1$ and any codimension \cite{casini2016g,casini2023entropic}, and $d=2,D=2$ \cite{casini2017relative}, $d=2,D=3$ \cite{casini2019irreversibility}, there are proofs using positivity of relative entropy between the UV fixed point vacuum state and the one of the theory with nontrivial RG flow. These in fact could be extended to $d=2$ and any codimension, because the relative entropy in question is dominated by the universal logarithmic term and positivity  gives the desired result.    

\begin{table}[h]
\begin{center}\begin{tabular}[c]{l| p{2.2cm}|p{2.2cm}|p{2.2cm}|p{2.2cm}|p{2.2cm}}
    \label{table2}
      $ _d\diagdown^D $  & $2$ & $3$ & $4$ & $5$ & $\cdots $\\
      \hline
      1 & \scriptsize{positivity of relative entropy}  &  \scriptsize{positivity of relative entropy} & \scriptsize{positivity of relative entropy} & \scriptsize{positivity of relative entropy} & \scriptsize{positivity of relative entropy}  \\
          \cline{2-6}
      2 & \scriptsize{SSA or positivity of relative entropy} & \scriptsize{SSA + QNEC or positivity of relative entropy} &  \scriptsize{SSA + QNEC or positivity of relative entropy} &  \scriptsize{SSA + QNEC or positivity of relative entropy} &  \scriptsize{SSA + QNEC or positivity of relative entropy}\\
     \cline{2-6}
      3 &  & \scriptsize{SSA} & \scriptsize{SSA + QNEC} & \scriptsize{SSA + QNEC} & \scriptsize{SSA + QNEC} \\
        \cline{3-6} 
      4 & \multicolumn{2}{c|}{}& \scriptsize{SSA} & \scriptsize{SSA + QNEC} & \scriptsize{SSA  + QNEC} \\
    \cline{4-6}
\end{tabular}\end{center}
\caption{Entropic proofs of defect irreversibility inequalities. SSA refers to strong subadditivity of entropy, and QNEC to quantum null energy condition (present paper).}
\end{table}

In this paper we extend these entropic proofs to the cases of $d=3,4$ (and the proof also covers $d=2$) and general $D$. This has the virtue of unifying all known defect irreversibility theorems, and extending them to the case of $d=3$ which has been refractory to treatment by other methods. The main idea is to generalize the use of SSA in the entropic proofs for $D=d$ extending it to higher codimension. In the rest of this introduction, we will summarize the central results of our work.

\subsection{Summary of main results and organization of the paper}

Our proof of defect irreversibility inequalities for $d \le 4$ and general $D$ involves two key steps.
The first is that the quantity that plays the role of the entropy in those theorems is now the relative entropy between the vacuum (of the theory undergoing the nontrivial RG flow) and the UV fixed point vacuum, reduced to a sphere, and compared at the null cone.  The second key idea is to use the quantum null energy condition (QNEC) \cite{bousso2016quantum}, which  involves the expectation value of the stress tensor. This is consistent with, and gives an explanation for, the appearance of an energy term in the universal quantity for codimension $D-d>1$. 

\begin{figure}[h]       
  \includegraphics[scale=0.45]{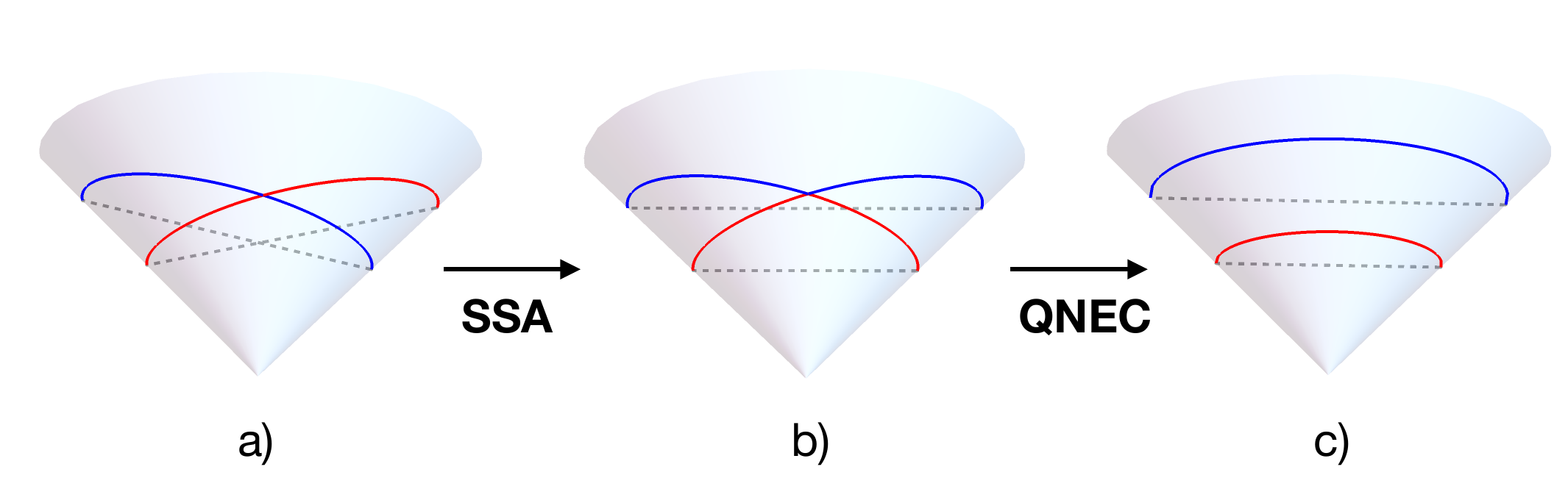}
      \caption{Steps for the defect irreversibility proof. a) Boosted spheres centered around the defect. b) SSA inequality with unions and intersections of boosted spheres. c) Replacement of unions and intersections by horizontal spheres, using the QNEC.}
    \label{fig:unamasa}
\end{figure}

In simple terms, the idea of the proof is summarized by Fig. \ref{fig:unamasa}. The grey lines show the strong subadditivity inequality between boosted spheres on the light cone as it would be the case if there would be no bulk dimensions. This is known to give the irreversibility theorems for $d=2,3,4$ and no extra bulk dimensions. In this case the intersections and unions of the (rotationally symmetrized) boosted spheres give also spheres. This gives an inequality for the universal terms, producing the irreversibility theorems.  Now, when there is a bulk, the intersections and unions of the bulk spheres fail to give bulk regions corresponding to boundary spheres (left and middle panels of Fig. \ref{fig:unamasa}). To correct this it is convenient to rewrite the inequalities involving the relative entropy, and use the QNEC inequality  in the bulk. This is shown in the right panel. Then, using the QNEC in the bulk we obtain the same geometric inequality in terms of spheres that we would have obtained if the theory would have no bulk.     

Our main result is contained in the formula for the relative entropy on a sphere of radius $R$,
\be\label{eq:supremo}
R \, S_{\rm rel}''(R) - (d-3) S_{\rm rel}'(R) \ge 0\,.
\ee
This is independent of $D$, and will be shown to give rise to the defect irreversibility inequalities for $d=2, 3, 4$. For $D=d$, it reduces to the analog formula for QFTs without defects obtained in \cite{casini2017markov}.
 The picture that emerges is that of inequalities for the relative entropy of increasing order in derivatives, capable of reaching to the universal terms for higher dimension. For $d=1,2$ it is enough with positivity, and $d=3,4$ requires the QNEC,  that gives the inequality  (\ref{eq:supremo}) with second derivatives of the relative entropy with respect to the sphere radius. It is plausible that inequalities involving two more derivatives would give the cases $d=5,6$, etc. However, no inequality of this type is known at present.    

 Establishing these results will require a technical development that is also interesting on its own sake. This involves the use of infinitesimal deformations of null boundaries on the light-cone, and their combination with restrictions from Lorentz invariance. This is the subject of Sec. \ref{sec:irrev}, where we provide a simple derivation of (\ref{eq:supremo}) when there are no defects. A second technical point involves an analysis of the QNEC on the light-cone for a CFT (instead of on the null plane for a QFT, as is usually done). This was studied in \cite{koeller2016holographic} and here we re-derive this result in Sec. \ref{sec:qnec}. As we discussed before, this will play a central role in obtaining the irreversibility theorems.

The remaining sections \ref{sec:defects} -- \ref{sec:holography} are devoted to theories with defects. In Sec. \ref{sec:defects} we discuss the setup; we study the entanglement and relative entropy and the contribution of the modular Hamiltonian on the lightcone Cauchy surface.  Sec. \ref{sec:defectirrev} contains our main result: the proof of  (\ref{eq:supremo}) using infinitesimal deformations and the QNEC. Consequences of this formula for defect irreversibility are also studied. In Sec. \ref{sec:another}  we discuss the proof making use of finite deformations, along the lines of Fig. \ref{fig:unamasa}. Finally, in Sec. \ref{sec:holography} we make some comments on the relation between our results and holographic RG flows where quantum effects give the leading contribution.  This understanding reveals that our proof also holds for a generalization of defect irreversibility where a defect is placed in the AdS boundary and  where the bulk corresponds to any QFT in AdS, rather than a CFT in Minkowski space.

\section{Irreversibility theorems from infinitesimal deformations}
\label{sec:irrev}

In this section we revisit the irreversibility theorems in general $d$-dimensional Minkowski space-time without defects. We simplify the geometry of the proofs considering only infinitesimal deformations and the relevant variation kernels of the entropy. This will be convenient for the generalizations to include the presence of defects in the following sections.

For a $d$-dimensional  CFT without defects the vacuum EE on a sphere has the form
\be\label{espa}
S(R)=\mu_{d-2}\,R^{d-2}+\mu_{d-4}\, R^{d-4}+\ldots+ \left\lbrace \begin{array}{ll} (-)^{\frac{d-2}{2}} 4\,A\, \log(R/\epsilon)\,& d\;  \textrm{even}\,.\\ (-)^{\frac{d-1}{2}} F\,& d\,\,\textrm{odd} \,. \end{array}\right.
\ee
For a CFT the coefficients $\mu_{k}\sim \epsilon^{-k}$ are proportional to inverse powers of the cutoff. We are interested in unitary Lorentz invariant QFTs obtained from renormalization group flows of CFT UV fixed points. In terms of the EE, the UV fixed point is approached for $R \to 0$, while the IR fixed point corresponds to $R \to \infty$. In this latter case, we have the same expression (\ref{espa}), but the coefficients get corrected by terms involving powers of physical scales. The last term is the one exhibiting the dimensionless RG charges. The terms involving $A$, for $d$ even, and $F$, for odd $d$, can be computed as the universal terms of the free energy of  the CFT in a $d$-dimensional Euclidean sphere. $A$ is also     
 the coefficient of the Euler term in the trace anomaly in even dimensions. The program of establishing the irreversibility of renormalization group flows involves proving that the universal quantities decrease between UV and IR fixed points.
 
If we subtract the entropy $S(R)$ from the corresponding function of the UV fixed point we get a finite quantity  $\Delta S(R)$ that vanishes for $R\rightarrow 0$ (see discussion in Subsec. \ref{subsec:newproof}). The irreversibility inequality is \cite{casini2017markov}
 \be
R \,\Delta S''(R)-(d-3)\, \Delta S'(R)\le 0\,. \label{ineirre}
\ee
 For $d=2,3,4$ it proves the monotonicity of the dimensionless RG charges.  For higher dimensions it implies $\Delta \mu_{d-2}\le 0$ and $\Delta \mu_{d-4}\ge 0$ for the change in the coefficients between the IR and UV fix points, though it is not enough to say anything about the dimensionless coefficients. The proof of (\ref{ineirre}) uses the Markov property of the CFT vacuum \cite{casini2017modular} and strong subadditivity (SSA) applied to multiple boosted spheres. Our goal in this section is to simplify this proof, exhibiting the combined effects of Lorentz invariance and strong subadditivity in a much simpler way.

\subsection{Infinitesimal null deformations and Lorentz invariance}\label{subsec:lorentz}

We consider surfaces on the null cone determined by the equation
\be
x^0=|\vec{x}|=\gamma(\hat{x})\,,
\ee
where $\hat{x}=\vec{x}/|\vec{x}|$ gives the angular directions on the cone. 
We Taylor expand the entropy for surfaces on the null cone around some surface $\gamma(\hat{x})$ that we will later take to be a sphere of radius $R$:
\be
S(\gamma+\delta \gamma)=S(\gamma)+\int d\Omega_1\, \frac{\delta S}{\delta\gamma(\hat{x}_1)}\, \delta \gamma(\hat{x}_1)+\frac{1}{2} \int d\Omega_1\, d\Omega_2\, \frac{\delta^2 S}{\delta \gamma(\hat{x}_1)\, \delta \gamma(\hat{x}_2)}\,  \delta \gamma(\hat{x}_1)\, \delta \gamma(\hat{x}_2)+\cdots\label{entro}
\ee
The integrals are over the normalized volume element of the unit sphere, $\int d\Omega=1$.  
The off diagonal part of the kernel in the second order term must be negative by SSA,
\be
\frac{\delta^2 S}{\delta \gamma(\hat{x}_1)\, \delta \gamma(\hat{x}_2)}\le 0\,,\hspace{1.3cm} \hat{x}_1 \neq \hat{x}_2\,.
\ee

However, there are also localized delta functions, and delta function derivatives, in this kernel. For the vacuum conformal case these singular local terms are the only terms and the non local part of the kernel vanishes. This is due to the Markov property of the vacuum that implies the entropy is a local functional on the cone for a CFT.  The conformal kernel  includes $d-2$ derivatives of the delta function in $d$ and $d+1$ dimensions, taking $d$ even \cite{casini2018all}. For a QFT the singular part of the kernel consists of the one of the UV fixed point plus possible delta functions or derivatives with coefficients proportional to some UV relevant coupling constants and diverging with powers of the cutoff. All divergent terms in the entropy are also Markovian,  in the sense that are local and cancel out in the SSA inequality.\footnote{This simplicity of the singular structure only holds for perturbations on the light cone where the Markov property of a CFT implies distributions supported at a point. For spatial perturbations other singular distributional kernels appear \cite{mezei2015entanglement,faulkner2016shape}.}     

For perturbations around a sphere of radius $R$ we write these kernels 
\bea 
S_1(\hat{x_1},R) &=& S_1(R)=\left.\frac{\delta  S}{\delta\gamma(\hat{x}_1)}\right|_{\gamma=R}\,,\label{a}\\
S_{12}(\hat{x}_1,\hat{x}_2,R)&=&S_{12}(\hat{x}_1\cdot \hat{x}_2,R)=\left.\frac{\delta^2  S}{\delta \gamma(\hat{x}_1)\, \delta \gamma(\hat{x}_2)}\right|_{\gamma=R}\,.\label{b}
\eea
In the first equations on each line we used rotational symmetry on the sphere. 
If we specialize (\ref{entro}) for the case of constant $\delta \gamma=dR$ around $\gamma=R$ we get
\begin{eqnarray}
 S'(R)&=&\int d\Omega_1\, \left.\frac{\delta  S}{\delta\gamma(\hat{x}_1)}\right|_{\gamma=R}=S_1(R)\,,\label{ddx}\\
 S''(R)&=&\int d\Omega_1\, d\Omega_2\, \left.\frac{\delta^2  S}{\delta \gamma(\hat{x}_1)\, \delta \gamma(\hat{x}_2)}\right|_{\gamma=R}=\int d\Omega_1\, d\Omega_2\, S_{12}(\hat{x}_1\cdot \hat{x}_2,R)\,.
  \label{dots}
\end{eqnarray}

Let us analyse the implications of Lorentz invariance. Boosting the horizontal sphere of radius $R$ in the direction $\hat{n}$, the new $\gamma$ will have a maximum $ \lambda\, R$, and a minimum $R/\lambda$, with $\lambda>1$, in the direction $\hat{n}$. The equation is
\be\label{eq:boost}
\gamma_{\lambda}(\hat{x})= \frac{2 R}{\lambda+\lambda^{-1}-(\lambda-\lambda^{-1})\, (\hat{x} \cdot \hat{n})}\,.
\ee
For any $\lambda$, the Lorentz invariance of the vacuum entanglement entropy is expressed in the form
\begin{equation}
\label{Lorentz}
S(\gamma_\lambda) = S(\gamma_{\lambda=1})\,.
\end{equation} 
Expanding (\ref{eq:boost}) in the boost parameter $d\lambda = \lambda - 1$ we obtain
\begin{equation}
\label{expand}
\frac{\delta \gamma(\vec{x})}{R} = (\hat{x} \cdot \hat{n})\,  d\lambda   + \left( (\hat{x} \cdot \hat{n})^2 -\frac{(\hat{x} \cdot \hat{n})}{2}-\frac{1}{2} \right) d\lambda^2 + \cdots
\end{equation}

Combining this expansion with (\ref{Lorentz}) and the expansion for the entropy (\ref{entro}), Lorentz invariance implies, for each order in $d\lambda$, a relation for the kernels of the entropy expansion. 
For example, at first order in $d \lambda$ we simply obtain from (\ref{Lorentz})
\begin{equation}
\int d\Omega_1\,  \,(\hat{x}_1\cdot \hat{n})\,S_1(R)=0\, ,
\end{equation}
which is certainly true.
At order $d \lambda^2$ we obtain, using (\ref{ddx}),
\begin{equation}
\label{order2}
\int d\Omega_1\,   ((\hat{x}_1\cdot \hat{n})-1)(2 (\hat{x}_1\cdot \hat{n})+1) \,S'(R) + R\,\int d\Omega_1\,d\Omega_2\,   (\hat{x}_1\cdot \hat{n}) \,(\hat{x}_2\cdot \hat{n}) \,S_{12}(\hat{x}_1\cdot \hat{x}_2) = 0\,.
\end{equation}

The first term can be evaluated using
\be
\int d\Omega_1\, \hat{x}_1^i\, \hat{x}_1^j=  (d-1)^{-1}\, \delta^{ij}\,.
\ee
For evaluating the second term note that by rotational invariance
\be
\int d\Omega_1\,d\Omega_2\, \,S_{12}(\hat{x}_1 \cdot \hat{x}_2)\,\hat{x}_1^i \hat{x}_2^j= \frac{\alpha(R)}{d-1}\,\delta^{ij}\,,
\ee
where
\be\label{eq:2pointsym}
\alpha(R)=\int d\Omega_1\,d\Omega_2\,\,S_{12}(\hat{x}_1 \cdot \hat{x}_2)\,(\hat{x}_1\cdot \hat{x}_2)\,.
\ee
Then, from Lorentz invariance, eq. (\ref{order2}), we get
\be
R\,\alpha(R)-(d-3)\, S'(R)=0\,.\label{eqq}
\ee

\subsection{Proof of irreversibility formula}\label{subsec:newproof}

Now we consider subtracting the entropy $S_\sigma$ of the UV CFT fixed point vacuum state $\sigma$ from the entropy of the theory
\be
\Delta S=S-S_\sigma. 
\ee
 This operation is meant to eliminate all contact terms in the kernel $S_{12}$. Massive divergent contributions are also subtracted. Divergent contributions are local and Markovian, and do not affect the kernel at non-coinciding points. Therefore the kernel $\Delta S_{12}$ coincides with $S_{12}$ at non-coinciding points and does not contain singular terms. By strong subadditivity we then have $\Delta S_{12} \le 0$. 
Since $ (1-\hat{x}_1 \cdot \hat{x}_2) \ge 0$ we get the inequality\footnote{For $d\le 3$ there are no second derivatives of the delta function or higher singular kernels, so the pre-factor in the integrand that vanishes at coincident points is enough to eliminate the contact terms, and the subtraction of the UV entropy is not necessary.}
\be
\int d\Omega_1\,d\Omega_2\,(1-\hat{x}_1\cdot \hat{x}_2)\,\Delta S_{12}(\hat{x}_1 \cdot \hat{x}_2)=\Delta S''(R)-\Delta \alpha(R)\le 0\,,
\ee
where we have used (\ref{dots}).
Replacing the second term here using the subtracted expression corresponding to (\ref{eqq}) obtains
\be\label{theinequality}
 R \,\Delta S''(R) - (d-3) \Delta S'(R) \le 0\,.
\ee

We used Lorentz and rotational invariance to translate the strong subadditivity inequality  $\Delta S_{12}\le 0$ (which makes crucial use of the Markov property on the null cone) into an inequality, (\ref{theinequality}), in terms of the derivatives with respect to $R$ of the subtracted entanglement entropy of a sphere of radius $R$. The area-terms and sub-area-term theorems in all dimensions follow from this inequality \cite{casini2017markov}. These coincide with the $C$, $F$, $A$ theorems in $d=2,3,4$, respectively. 

For applications to QFT with defects below, it will be convenient to recast this result in terms of the relative entropy. Calling $\sigma$ to the vacuum state of the CFT ultraviolet fixed point and $\rho$ to the vacuum of the QFT, we compute the relative entropy 
\be
S_{\rm rel}\equiv S(\rho|\sigma)= \Delta \langle H_\sigma \rangle- \Delta S\,,\label{rel}
\ee
where 
\be
\Delta \langle H_\sigma \rangle=\langle H_\sigma\rangle_\rho-\langle H_\sigma \rangle_\sigma\,,
\ee
and $H_\sigma$ is the modular Hamiltonian corresponding to $\sigma$. The modular Hamiltonian of a Markovian state also satisfies the Markov property, i.e. the SSA combination of modular Hamiltonians cancels at the operator level \cite{petz2007quantum},
\be
S(A)+S(B)-S(A\cup B)-S(A\cap B)=0\leftrightarrow H(A)+H(B)-H(A\cup B)-H(A\cap B)=0\,.
\ee
As the CFT vacuum $\sigma$ is Markovian on the null cone, the contributions of the modular Hamiltonian on the relative entropy will cancel when using the SSA combination.  As a result  eq. (\ref{theinequality}) holds as well for the relative entropies, with just a change of sign
\be
R \, S_{\rm rel}''(R) - (d-3) S_{\rm rel}'(R) \ge 0\,. \label{pppp}
\ee

We remark that
the relative entropy (\ref{rel}) between vacuuum states of theories at different scales is a delicate quantity though. Subtleties --- dependences on the Cauchy surface where the states are compared --- and divergences, were studied in \cite{casini2017relative}. The upshot is that the inequality (\ref{pppp}) makes perfect sense when evaluated on the null cone.   

\section{QNEC on the light-cone}\label{sec:qnec}

In this section we change subject: we study the quantum null energy condition (QNEC) on the light-cone for a CFT.  This has been previously analyzed in \cite{koeller2016holographic}. Here our motivation is two-fold. First, we will clarify and explain a few points related to the QNEC, and its application on the null cone. And second, the formula we will obtain will play a central role in our irreversibility proof for theories with defects.

The QNEC was discovered taking the flat space limit of the quantum focusing conjecture \cite{bousso2016quantum}, a proposed quantum generalization of the null energy condition in gravity. Direct proofs in flat-space QFT for free fields \cite{bousso2016proof}, and interacting fields \cite{balakrishnan2019general,balakrishnan2022entropy}, have been produced (see also the holographic proofs \cite{koeller2016holographic,leichenauer2018energy}).

\subsection{The null plane}

Consider first the null plane $x^-=0$ in Minkowski space, with $x^\pm=x^0\pm x^1$, and call $\vec{y}=(x^2,\cdots,x^d)$ to the vector of coordinates on the plane that are transversal to the null line. A $(d-2)$-dimensional surface $\gamma$ on the null plane is determined by the equations 
\be
x^-=0\,, \,x^+=\gamma(\vec{y})\,.
\ee  
Let us consider a one parameter deformation $\gamma_s$ of $\gamma$ along the null plane, with $\gamma_0=\gamma$,  such that for each fixed $\vec{y}$ the parameter $s$ is an affine parameter along the null line corresponding to $\vec{y}$. That is,
\be
\gamma_s(\vec{y})= a(\vec{y})\, s+ \gamma(\vec{y})\,.
\ee
It is also required that the deformation be towards one side of $\gamma$, namely $a(\vec{y})\ge 0$. 
The QNEC is the inequality
\be
 2 \pi\,\int_\gamma \, \left(\frac{d\gamma_s}{ds}\right)^2\, \langle T_{++} \rangle 
\ge \left.\frac{d^2  S(\gamma_s)}{ds^2}\right|_{s=0}\,.\label{hh}
\ee
This is valid for any state $\rho$. On the right hand side $S(\gamma_s)$ is the entanglement entropy of this state across $\gamma_s$.

The inequality then gives lower bounds for the energy density in terms of a second derivative of the entanglement entropy. For our purposes it is convenient to use the inequality expressed in terms of the relative entropy. For this we want to subtract vacuum quantities to the inequality. Because of the boost symmetry, the vacuum entropy $S_{\sigma}(\gamma)$ is independent of the surface $\gamma$ on the null plane \cite{casini2018all}. In addition, the vacuum modular Hamiltonian of the region $\gamma$ written on the null plane has a universal expression in terms of the stress tensor  \cite{casini2017modular,wall2012proof},
\be
H_\gamma= 2 \pi\,\int_{x^+>\gamma(\vec{y}),x^-=0} d^{d-2}y\,dx^+\, (x^+-\gamma(\vec{y}))\, T_{++}\,.\label{valid}
\ee
Using the expression  
\be
S_{\rm rel}(\gamma)\equiv S_\gamma(\rho|\sigma)=\langle H_\gamma\rangle_\rho-\langle H_\gamma\rangle_{\sigma}-( S_\rho(\gamma)-S_{\sigma}(\gamma))\,\label{ee}\ee 
for the relative between the state $\rho$ and the vacuum $\sigma$, and that $\langle T_{++}\rangle_{\sigma}=0$,
 the QNEC writes, more compactly, 
\be
\frac{d^2 S_{\rm rel}(\gamma_s)}{ds^2}\ge 0\,. \label{ff}
\ee
This is a convexity property of the relative entropy with respect to arbitrary affine parameter deformations on the null plane.\footnote{See \cite{Longo:2018obd} for an interesting identification of this inequality as the second derivative with respect to the parameter of modular translations corresponding to half sided modular inclusions. Arbitrary affine parameter deformations on the null plane are modular translations \cite{casini2017modular}. It is also interesting to note that, due to monotonicity of relative entropy, the inequality is still satisfied for non affine parametrizations of the form $\gamma_s=\gamma + a\, s+ b\, s^2+\cdots$ provided $a\ge 0$ and $b\le 0$.}

This second derivative can also be written as
\be
\frac{d^2 S_{\rm rel}(\gamma_s)}{ds^2}=\int d^{d-2}y_1\, d^{d-2}y_2\, \frac{\delta^2 S_{\rm rel}(\gamma)}{\delta \gamma(\vec{y}_1)\,\delta \gamma(\vec{y}_2)}\,\frac{d\gamma(\vec{y_1})}{ds}\,\frac{d\gamma(\vec{y_2})}{ds}\,.
\ee
Let us write the kernel as a sum of a smooth part $k(\vec{y}_1,\vec{y}_2)$ and a local part,\footnote{ Because the relative entropy contains the subtracted entropy and is a function of regions on the null plane this is the only possible singular structure of the kernel.}
\be
\frac{\delta^2 S_{\rm rel}(\gamma)}{\delta \gamma(\vec{y}_1)\,\delta \gamma(\vec{y}_2)}=k(\vec{y}_1,\vec{y}_2)+ \delta(\vec{y}_1-\vec{y}_2)\frac{d}{d \gamma(\vec{y})}\frac{\delta S_{\rm rel}(\gamma)}{\delta \gamma(\vec{y})}\,.
\ee
 For the smooth part,  when the relative entropy is decomposed as in (\ref{ee}), the contributions of the modular Hamiltonian and of the entropy $S_\sigma(\gamma)$ vanish because of the Markov property of the vacuum \cite{casini2017modular}. The contribution of $-S_\rho(\gamma)$ is automatically positive because of strong subadditivity, and this does not require affine parametrization. Then $k(\vec{y}_1,\vec{y}_2)\ge 0$. Hence, a good deal of what is involved in the inequality is the SSA inequality. The other information involved in the inequality is the so called local QNEC,
\be
\frac{d}{d \gamma(\vec{y})}\frac{\delta S_{\rm rel}(\gamma)}{\delta \gamma(\vec{y})}\ge 0\,.\label{lqnec}
\ee
This has been shown to be positive for UV free models in $d>2$ \cite{bousso2016proof} and to saturate for interacting models \cite{balakrishnan2022entropy}. 

\subsection{The null cone}

The QNEC  holds for any QFT. For a CFT  there are conformal mappings carrying the null plane into the null cone, and keeping the vacuum state invariant. Therefore, we get the same inequality where now the null deformations along the null cone correspond to the transformation of the affine parameter on the null line that follow from the conformal mapping. We take the tip of the null cone at the origin of coordinates, and map the null plane to the future null cone. We take $|\vec{x}|=x^0$ as affine parameter along a null line of the cone. A null line through any fixed $\vec{y}$ is mapped to an angle direction $\hat{x}$ on the cone. We get a general relation between affine parameters of the form
\be
|\vec{x}|= \frac{C_1(\vec{y})}{C_2(\vec{y})+ x^+}\,,
\ee
where $C_1>0$, and the map is limited to the range $x^++C_2>0$.

Defining a surface $\gamma$ on the null cone by the equation $|\vec{x}|=\gamma(\hat{x})$, the allowed deformations are of the form
\be
\gamma_s(\hat{x})=\frac{\gamma(\hat{x})}{1+s \, a(\hat{x})}\sim \gamma(\hat{x})\,(1-s \,a(\hat{x})+ s^2\, a(\hat{x})^2+\cdots)\,,\hspace{.7cm}a(\hat{x})\ge 0\,.
\ee
For these deformations we have the QNEC on the null cone
\be
\frac{d^2 S_{\rm rel}(\gamma)}{ds^2}\ge 0\,.
\ee
The non local part of the QNEC is positive because of SSA and the Markov property, as above, provided the linear deformations for different points have the same sign.
Changing variables, the inequality (\ref{lqnec}) gives the local QNEC on the cone
\be
\left(\frac{d}{d \gamma(\hat{x})}\frac{\delta S_{\rm rel}(\gamma)}{\delta \gamma(\hat{x})}+\frac{2}{\gamma(\hat{x}) }\,\frac{\delta S_{\rm rel}(\gamma)}{\delta \gamma(\hat{x})}\right)\ge 0\,. \label{mir}
\ee
Combining this with the positivity of the second variation kernel at non coinciding points, we get
\be
 \frac{\delta^2 S_{\rm rel}}{\delta \gamma(\hat{x}_1)\, \delta \gamma(\hat{x}_2)}+\delta(\Omega_1-\Omega_2)\, \frac{2}{\gamma(\hat{x}_1)}\,  \frac{\delta S_{\rm rel}}{\delta\gamma(\hat{x}_1)}\ge 0\,,\label{ddss}
\ee
in the sense that this kernel gives a positive number upon integration with any positive function of the two directions on the sphere.

\section{Defects and RG charges}\label{sec:defects}

We now move to the  main topic of this paper: renormalization group flows on defects, embedded in bulk CFTs. The results developed in Secs. \ref{sec:irrev} and \ref{sec:qnec} will play a key role in our proof of the irreversibility of such flows. In this section we will review the basic aspects of the setup. The irreversibility formula for defects will be derived in Sec. \ref{sec:defectirrev}, and a different (but equivalent) analysis will be presented in Sec. \ref{sec:another}.

\subsection{Entanglement and relative entropies}\label{subsec:EEdefect}

Let the CFT live in $D$ space-time dimensions and the defect in a planar static $d$-dimensional hypersurface passing through the origin. Let $\sigma$ be the vacuum of the theory with the conformal defect. The defect RG flow is started by a relevant operator (or combination thereof) on the defect action
\be
S=S_0+ g\, \int d^dx\, {\cal O_\Delta}\,, 
\ee
where the conformal dimension $\Delta < d$. We call $\rho$ to the vacuum state of this new theory.

A  quantity of interest is the entropy of a sphere centered at the origin. For a conformal defect we have an expansion analogous to (\ref{espa}): 
\bea
S(R)&=&\mu_{D-2}\,R^{D-2}+\mu_{D-4}\, R^{D-4}+\ldots+\tilde{\mu}_{d-2}\,R^{d-2}+\tilde{\mu}_{d-4}\, R^{d-4}+\ldots \nonumber \\
&+&\left\lbrace \begin{array}{ll} (-)^{\frac{D-2}{2}} 4\,A\, \log(R/\epsilon)\,& D\;  \textrm{even}\\ (-)^{\frac{D-1}{2}} F\,& D\,\,\textrm{odd}  \end{array}\right. +\left\lbrace \begin{array}{ll} (-)^{\frac{d-2}{2}} 4\,\tilde{A}\, \log(R/\epsilon)\,& d\;  \textrm{even}\,.\\ (-)^{\frac{d-1}{2}} \tilde{F}\,& d\,\,\textrm{odd} \,. \end{array}\right.
\eea
Here quantities with tilde correspond to contributions of the defect. We will denote the (last) universal terms in this expression by $S_\text{univ}$. For a RG flow on the defect, the defect contribution will change with $R$ producing different coefficients at the UV and IR fixed points. On the other hand, since we have a bulk CFT, the contribution of the bulk remains fixed along the RG. Thus, if we again subtract the entropy of the UV fixed point to $S(R)$ we get a difference $\Delta S(R)$ that is insensitive to the bulk $D$-dimensional terms. We have $\lim_{R\rightarrow 0}\Delta S(R)=0$ and
\be\label{eq:DeltaSlargeR}
\lim_{R\rightarrow \infty}\Delta S(R)=\Delta \tilde{\mu}_{d-2}\,R^{d-2}+\Delta \tilde{\mu}_{d-4}\, R^{d-4}+\ldots+ \left\lbrace \begin{array}{ll} (-)^{\frac{d-2}{2}} 4\,\Delta \tilde{A}\, \log(R/\epsilon)\,& d\;  \textrm{even}\,.\\ (-)^{\frac{d-1}{2}} \Delta \tilde{F}\,& d\,\,\textrm{odd} \,. \end{array}\right.
\ee

This may suggest that the relevant RG charges for a defect flow are the $\tilde{F}, \tilde{A}$ associated to the defect. In fact, if we could show again inequality (\ref{ineirre}) it would imply the monotonicity of the defect entropy universal coefficients for $d\le 4$. However, as we will see, it is not possible to prove this inequality from SSA essentially because rotational and boost symmetries are now restricted to directions parallel to the defect.    

In fact, explicit examples and other considerations prompted \cite{Jensen:2018rxu, kobayashi2019towards} to conjecture that the RG charge should contain along with the universal entropy coefficient a coefficient associated to the energy produced by the defect. The idea is that the RG charge should correspond to the universal part  induced by the defect on the free energy on the $D$-dimensional Euclidean sphere, as was the case of the RG without bulk. This sphere now contains a defect included in the equator and the free energy corresponds to the logarithm of the partition function plus a energy term that does contribute to the universal part. This contribution vanishes without the defect \cite{kobayashi2019towards}.   

Our analysis will be in real time rather than in imaginary time, and in our case the energy contribution comes from considering a relative entropy rather than the entropy. The connection with the Euclidean understanding comes from the fact that the relative entropy between a state and a thermal equilibrium state is the difference of the free energies between the two states. The ``thermal equilibrium state'' corresponds to the UV fixed  point state $ \sigma$ and the other state is the defect flow state $\rho$. 

Let us describe this relative entropy in detail.  Since the two models have different Hamiltonian we have to fix a Cauchy surface to compare the states. In other words the relative entropy will depend on the choice of this surface. We choose to compare them on the null cone \cite{Casini:2016fgb,casini2017relative, casini2019irreversibility, casini2023entropic}. Another point of view is to quench $g\rightarrow 0$ as the defect hits the null cone. This leaves us with a state $\rho$ that now propagates with the unperturbed Hamiltonian on the future null cone, and the relative entropy will be Cauchy surface independent. 

The relative entropy is
\be\label{eq:Srel1}
S(\rho|\sigma)\equiv S_{\rm rel}(R)= \textrm{tr} (\rho \, \log \rho- \rho \, \log \sigma)= \Delta \langle H \rangle-\Delta S\,.   
\ee
Here $H=-\log \sigma$ is the modular Hamiltonian corresponding to $\sigma$, and $\Delta \langle H \rangle=\langle H\rangle_\rho- \langle H\rangle_\sigma$. 
The modular Hamiltonian has a universal expression proportional to the stress tensor.  $\Delta \langle H \rangle$ generically depends on the Cauchy surface where we quantize the theory. We will consider our theory to be quantized in the null cone, where the divergences of the modular Hamiltonian match the divergences of the entropy \cite{casini2017relative}.  Let us discuss what we obtain in this limit.

\subsection{Defect stress tensor}

In order to evaluate $\Delta \langle H \rangle$ in (\ref{eq:Srel1}), we will need the expectation value of the stress tensor in the presence of the defect with nontrivial RG flow.

We split the coordinates as
\be
x_\mu=(x_\alpha, y_a)\;,\;\mu=0,\ldots, D-1\;;\;\alpha\;=0,\ldots, d-1\;;\; a=d,\ldots, D-1\,.
\ee
The defect is placed at
\be
y_a=0\,.
\ee
Operatorially, $T_{\mu\nu}$ is conserved and traceless away from the defect. Furthermore, by rotational invariance it can only depend on $y= (y_a y_a)^{1/2}$. The conservation condition $
\partial_a T^{a\alpha}=0$ requires
\be
\Delta \langle T_{a \alpha} \rangle =0\,.
\ee 
The remaining nonzero components can then be parametrized as 
\bea\label{eq:paramT}
\Delta \langle T_{\alpha \beta}\rangle&=& h(y) \eta_{\alpha \beta}\\
\Delta \langle T_{ab}\rangle&=& f_1(y) \delta_{ab}+f_2(y) \left(\frac{y_a y_b}{y^2}-\frac{\delta_{ab}}{D-d} \right)\,. \nonumber
\eea
Here we used Poincar\'e invariance along the defect to constrain $T_{\alpha \beta}$, and rotational invariance in the transverse directions to fix $T_{ab}$. 

Requiring vanishing trace relates
\be\label{eq:f1}
f_1(y)=-\frac{d}{D-d} h(y)\,.
\ee
The conservation condition $\partial_a T^{ab}=0$ gives, on the other hand,
\be\label{eq:f2}
\frac{D-d-1}{D-d} \left(f_2'(y)+\frac{D-d}{y} f_2(y) \right)+f_1'(y)=0\,.
\ee
For $D=d+1$ this leads to the vanishing of $\langle T_{\alpha\beta}\rangle$. More generally, we have an arbitrary free function, which we may take as $h(y)$ by doing the integration 
\begin{equation}\label{eq.f2}
    f_2(y)=-\frac{d}{(D-d-1)y^{D-d}}\int_y^\infty du\, u^{D-d} h'(u)\,.
\end{equation}
The integration constant here is chosen in order to reproduce the correct conformal limit (as discussed next).

The form of the stress tensor is further determined if the  states are conformally invariant. Since there is no dimensionful coupling, and the stress tensor has dimension $D$, $h(y)= h_0/y^D$. As a result,
\bea\label{eq:T0}
\Delta \langle T_{\alpha \beta}\rangle &=&  \frac{h_0}{y^D}\,\eta_{\alpha \beta}\\
\Delta \langle T_{ab}\rangle &=& -\frac{d}{D-d}  \frac{h_0}{y^D}\,\delta_{ab}+\frac{D}{D-d-1}\frac{h_0}{y^D}\, \left(\frac{y_a y_b}{y^2}-\frac{\delta_{ab}}{D-d} \right)\nonumber\\
\Delta \langle T_{a \alpha} \rangle &=&0\,. \nonumber
\eea
Only the constant $h_0$ is arbitrary, and this depends on the type of conformal defect. We get to the expression (\ref{eq:T0}) when compute the stress tensor at large distances from the defect, where the IR fixed point has been reached. The quantity $h_0$ is then the difference between the IR and UV coefficients. To compare with \cite{kobayashi2019towards,billo2016defects}, 
\be
h_0=\frac{D-d-1}{D} \,\Delta a_T  \,.
\ee

\subsection{The modular Hamiltonian for defect RG flows}\label{subsec:DeltaH}

The modular Hamiltonian in the null limit reads 
\be
H=\frac{2\pi}{R} \int d^{D-2}\Omega\,\int_0^R\,d\lambda \,\lambda^{D-1}(R-\lambda) T_{\lambda \lambda}\,,
\ee
with $\lambda=\frac{x^0+r}2=r$, and 
\begin{equation}\label{eq:Tll}
    T_{ \lambda \lambda}= T_{00}+\frac{  x^i x^j}{r^2}T_{ij}\,.
\end{equation}
Here $r=\sqrt{x^2+y^2}$. In terms of the previous functions,
\begin{equation}
    \Delta\langle T_{\lambda \lambda}\rangle= \frac{1}{D-d}\frac{y^2}{r^2}\left[ -D\,h(y)+(D-d-1) f_2(y)  \right ]\,.
\end{equation}

In terms of spherical coordinates,
\be
\vec x= r \cos \theta\,\hat n\;,\;\vec y= r \sin \theta\,\hat n'
\ee
(with $\hat n$ and $\hat n'$ unit vectors in $\mathbb R^{d-1}$ and $\mathbb R^{D-d}$, respectively) we have
\bea
\Delta \langle H \rangle&=& \frac{2\pi}{R} \text{Vol}(S^{d-2}) \text{Vol}(S^{D-d-1})\,\int_0^{\pi/2} d\theta\, |\cos \theta|^{d-2} |\sin \theta|^{D-d-1}\nonumber\\
&&\times\int_0^R dr\,r^{D-1}(R-r) \Delta \langle T_{\lambda \lambda}(r,y) \rangle\,.
\eea
Changing variables to $x=r \cos \theta$ and $y = r \sin \theta$, this becomes
\bea\label{eq:DeltaHtemp1}
\Delta \langle H \rangle&=& \frac{2\pi}{(D-d)R} \text{Vol}(S^{d-2}) \text{Vol}(S^{D-d-1})\,\int_0^R dy \int_0^{\sqrt{R^2-y^2}}dx\, x^{d-2} y^{D-d+1} \nonumber\\
&&\times \frac{R-r}{r}\left[ -D\,h(y)+(D-d-1) f_2(y)  \right ]\,.
\eea

The integral over $x$ gives
\bea
P(y) &\equiv& \frac{1}{R} \int_0^{\sqrt{R^2-y^2}}dx\, x^{d-2} \frac{R-r}{r} \nonumber\\
&=&\frac{R^{d-2}}{d-1} \left(1-\frac{y^2}{R^2} \right)^{\frac{d-1}{2}} \left[-1+ \frac{R}{y}\;{}_2 F_1(\frac{1}{2},\frac{d-1}{2},\frac{d+1}{2},1-\frac{R^2}{y^2}) \right]\,.
\eea
The large $R$ expansion gives powers of the form $R^{d-2}, R^{d-4}$ and so on. As expected, the null limit has reduced the super-extensive growth of the modular Hamiltonian, giving instead a leading area contribution $R^{d-2}$.
We want to isolate the universal term. For this, let us assume that $d$ is not an even integer, and take $R \to \infty$. The appropriate large $R$ term comes from the contribution of the hypergeometric function with the appropriate power to cancel the $R$-dependence. This gives
\be
P(y)_{\rm univ}= \frac{\Gamma\left( 1-\frac{d}{2}\right) \Gamma \left(\frac{d+1}{2} \right)}{\sqrt{\pi}\,(d-1)} y^{d-2}\,.\label{exph}
\ee
Note that the expansion of the factor $\left(1-\frac{y^2}{R^2} \right)^{\frac{d-1}{2}}$ in $P(y)$ does not contribute to the universal term if $d$ is not an integer (as we assume for the present derivation).

With this result, let us return to the calculation of (\ref{eq:DeltaHtemp1}). Plugging the expression for $f_2$ and integrating by parts the $h'(u)$ term,
\be
\Delta \langle H \rangle =-2\pi \text{Vol}(S^{d-2}) \text{Vol}(S^{D-d-1})\,\int_0^R dy  P(y)\,y\left (  y^{D-d} h(y)-d  \int_y^\infty du \,h(u) u^{D-d-1} \right )\,.
\ee
A more convenient form for what follows is obtained by changing the integration order in the double integral,
\bea
\Delta \langle H \rangle &=&-2\pi \text{Vol}(S^{d-2}) \text{Vol}(S^{D-d-1})\,\Bigg[\int_0^R dy  P(y)y^{D-d+1}h(y) \\
&& \qquad -d \int_0^R du \,h(u) u^{D-d-1}\int_0^u dy  P(y)  y -d  \int_R^\infty du \,h(u) u^{D-d-1} \int_0^R dy P(y)  y \Bigg]\,.\label{po} \nonumber
\eea

Now we want to isolate the universal contribution to $\Delta \langle H \rangle$.
Replacing $P(y)\rightarrow P(y)_{\rm univ}$ the integral appears dominated by the large $y$ region. In this regime, each of the first two terms diverges logarithmically with $R$, but they cancel each other. The third term gives 
\be\label{eq:DeltaHuniv}
\Delta\langle H\rangle_{\rm univ}=-\frac{2 (D-d-1) \, \pi^{D/2+1}}{\sin(\pi d/2)\, D \, \Gamma(d/2+1) \, \Gamma((D-d)/2)} \, \Delta a_T\,.
\ee
Crucially, this turns out to a difference of universal quantities at fix points. Note that for $d=2n$ even this expression gives a simple pole at $d=2n$; this is the dimensional regularization version of a logarithmic divergence $\sim \log R$. Equivalently, when the dimension $d$ is an even integer the expansion of $P(y)$ gives a term proportional to $y^{d-2} \log(y/R)$ that follows from expanding (\ref{exph}) around even $d$. In this case the first two integrals in (\ref{po}) give a term logarithmically divergent with $R$ while the last term does not contribute to the logarithmic term. 

Finally, combining the universal contributions of (\ref{eq:DeltaHuniv}) and  $\Delta S$ to the relative entropy gives
\be
S_\text{rel}(R) \Big|_\text{univ}=\Delta\langle H\rangle_{\rm univ}-\Delta S_\text{univ}=F_\text{univ}^{UV}-F_\text{univ}^{IR}
\ee
with the universal quantity
\be
F_\text{univ}=-\frac{2 (D-d-1) \, \pi^{D/2+1}}{\sin(\pi d/2)\, D \, \Gamma(d/2+1) \, \Gamma((D-d)/2)} \, a_T - S_{\rm univ}\,.
\ee
This coincides with the euclidean free energy proposal of \cite{kobayashi2019towards}.
This will be the subject of our irreversibility inequalities for $d \le 4$ -- dimensional defects.

\section{Irreversibility of defect RG flows}
\label{sec:defectirrev}

In this section we will establish our central result: the formula
\be\label{eq:central}
R \, S_{\rm rel}''(R) - (d-3) S_{\rm rel}'(R) \ge 0\,, 
\ee
for defect RG flows. Recall that $d$ is the spacetime dimension of the planar static defect, which is embedded in a $D$-dimensional CFT.  Furthermore, $S_\text{rel}(R) = S_\text{rel}(\rho_R|\sigma_R)$, with $\sigma_R$ the vacuum reduced density matrix on a sphere of radius $R$ for the UV fixed point, and $\rho_R$ the analog object for the theory with defect RG flow.  It is implicit that in this relative entropy we compare states on the null cone. 
Our analysis follows similar steps to that in Sec. \ref{sec:irrev}, but out of the defect we will use the QNEC instead of SSA.\footnote{See \cite{mezei2019quantum} for an application of the QNEC to quenches. } 

\subsection{QNEC in the presence of defects}

The argument that the QNEC applies in the presence of the defect is as follows.
Consider a $D$-dimensional QFT with a defect located on a surface of dimension $d<D$ that is invariant under the boosts that keep a null plane invariant.  Eq. (\ref{hh}) is then still valid. The entropy of the vacuum state is still invariant under null deformations for regions in the null plane. The Markov property and the explicit form (\ref{valid}) of the modular Hamiltonian are also still valid. See \cite{casini2018all}. We also have $\langle T_{++}\rangle_\sigma=0$; this can be checked by plugging the conformal answer (see e.g. (\ref{eq:T0})) into (\ref{eq:Tll}). Therefore (\ref{ff}) follows in the same way as above. For $\sigma$ the vacuum state of a CFT with a conformal defect, this inequality can be transformed into the one for regions with boundaries on the null cone. In particular, this applies to a planar static defect that passes through the tip of the cone.

\subsection{Proof of the defect irreversibility formula}

Let us now analyze the infinitesimal variations of the relative entropy. We write $\hat{x}$ for the unit vector in $D-1$ dimensions, describing the angular variables on a $S^{D-2}$ sphere. We decompose it into its components parallel and orthogonal to the defect,
\be
\hat{x}=\hat{x}^\parallel+\hat{x}^\perp\,.
\ee
Again we take a sphere of radius $R$ on the null cone $x^0=R$. Now we only have rotational and boost invariance in the directions parallel to the defect and rotational invariance in the perpendicular coordinates.
The  Taylor expansion of the relative entropy is again
\be
S_{\rm rel}(\gamma+\delta \gamma)=S_{\rm rel}(\gamma)+\int d\Omega_1\, \frac{\delta S_{\rm rel}(\gamma)}{\delta\gamma(\hat{x}_1)}\, \delta \gamma(\hat{x}_1)+\frac{1}{2} \int d\Omega_1\, d\Omega_2\, \frac{\delta^2 S_{\rm rel}(\gamma)}{\delta \gamma(\hat{x}_1)\, \delta \gamma(\hat{x}_2)}\,  \delta \gamma(\hat{x}_1)\, \delta \gamma(\hat{x}_2)+\cdots\label{entro2}
\ee
The two kernels are connected by the QNEC inequality (\ref{ddss}).

For perturbations around a sphere of radius $R$ we write these kernels
\bea 
S_1(\hat{x}_1) &=&\left.\frac{\delta  S_{\rm rel}}{\delta\gamma(\hat{x}_1)}\right|_{\gamma=R}\,,\label{a1}\\
S_{12}(\hat{x}_1, \hat{x}_2)&=&=\left.\frac{\delta^2  S_{\rm rel}}{\delta \gamma(\hat{x}_1)\, \delta \gamma(\hat{x}_2)}\right|_{\gamma=R}\,.\label{b1}
\eea
We keep in mind that they also depend on $R$, and are rotationally symmetric in the parallel and orthogonal components, though we will not make this explicit with the notation.
Also, we have
\begin{eqnarray}
 S_{\rm rel}'(R)&=&\int d\Omega_1\,S_1(\hat{x}_1)\,,\\
 S''_{\rm rel}(R)&=&\int d\Omega_1\, d\Omega_2\, S_{12}(\hat{x}_1, \hat{x}_2)\,.
  \label{dots1}
\end{eqnarray}

 Lorentz invariance gives again (\ref{order2}), with $\hat{n}$ a unit vector parallel to the defect,
\begin{equation}
\label{order21}
\int d\Omega_1\,   ((\hat{x}_1\cdot \hat{n})-1)(2 (\hat{x}_1\cdot \hat{n})+1) \,\,S_1(\hat{x}_1) + R\,\int d\Omega_1\,d\Omega_2\,   (\hat{x}_1\cdot \hat{n}) \,(\hat{x}_2\cdot \hat{n}) \,S_{12}(\hat{x}_1, \hat{x}_2) = 0\,.
\end{equation}
Using the rotational symmetry of $S_1$
\be
\int d\Omega_1\, (\hat{x}_1^\parallel)^i\, (\hat{x}_1^\parallel)^j \,S_1(\hat{x}_1) = \frac{\delta^{ij}}{d-1}\, \int d\Omega_1\,|\hat{x}_1^\parallel|^2\,S_1(\hat{x}_1)= \beta\,\frac{\delta^{ij}}{d-1}  \,,
\ee
with
\be
\beta=\int d\Omega\,|\hat{x}_\parallel|^2\,\,S_1(\hat{x})\,.
\ee
Analogously, we need for evaluating the second term in (\ref{order21}),
\be
\int d\Omega_1\,d\Omega_2\, \,S_{12}(\hat{x}_1 , \hat{x}_2)\,(\hat{x}_1^\parallel)^i (\hat{x}_2^\parallel)^j= \alpha\, \frac{\delta^{ij}}{d-1}\,,
\ee
where
\be
\alpha=\int d\Omega_1\,d\Omega_2\,\,S_{12}(\hat{x}_1, \hat{x}_2)\,(\hat{x}^1_\parallel \cdot \hat{x}^2_\parallel)\,.
\ee
Then, from Lorentz invariance, eq. (\ref{order21}), we get
\be
R\,\alpha+2 \beta-(d-1)\, S'_{\rm rel}(R)=0\,.\label{eqq22}
\ee

Since $ (1-\hat{x}_1^\parallel \cdot \hat{x}_2^\parallel) \ge 0$, integrating its product with the QNEC inequality (\ref{ddss}) we get 
\bea
&&\hspace{-1cm}\int d\Omega_1\,d\Omega_2\,(1-\hat{x}_1^\parallel\cdot \hat{x}_2^\parallel)\,(R\, S_{12}(\hat{x}_1, \hat{x}_2) + 2\, \delta(\Omega_1-\Omega_2)\, S_1(\hat{x}_1))\nonumber \\
&&\hspace{5cm}=R \,S''_{\rm rel}(R)-R \,\alpha+2\, S_{\rm rel}'(R)-2\, \beta\ge 0\,.
\eea
Notice that on the defect the delta function term does not contribute because of the $(1-\hat{x}_1^\parallel\cdot \hat{x}_2^\parallel)$ factor. Then on the defect the inequality reduces to the SSA. This is important because we do not have the QNEC for deformations on the defect but only in the CFT bulk. 
Adding this to (\ref{eqq22}) we get the desired inequality
\be\label{t}
 R \, S''_{\rm rel}(R) - (d-3)  S'_{\rm rel}(R) \ge 0\,.
\ee
This is equal to (\ref{pppp}), but now applies to spheres in a CFT in $D$ dimensions with a defect passing through the origin. This completes our proof.
  
In contrast to the analysis in section \ref{sec:irrev}, here we used the QNEC instead of the SSA to obtain this inequality.  This is clear from the fact that in the present case the contribution of the stress tensor (through the modular Hamiltonian) to the relative entropy is in general non trivial.  However, even in codimension $D-d=1$, where the modular Hamiltonian contribution vanishes, it is necessary to invoke the QNEC in the bulk to get the inequality.

\subsection{Consequences}

We now make explicit the consequences of (\ref{t}) in different dimensionalities. Combining the contributions of the modular Hamiltonian and $\Delta S$ to the relative entropy gives, in the IR limit of large $R$,
\be\label{eq:Srelprime}
-\lim_{R\rightarrow \infty}S_\text{rel}(R)=\Delta \mu'_{d-2}\,R^{d-2}+\Delta \mu'_{d-4}\, R^{d-4}+\ldots+ \left\lbrace \begin{array}{ll} (-)^{\frac{d-2}{2}} 4\,\Delta A'\, \log(R/\epsilon)\,& d\;  \textrm{even}\,.\\ (-)^{\frac{d-1}{2}} \Delta F'\,& d\,\,\textrm{odd} \,. \end{array}\right.
\ee
We use quantities with primes in order to distinguish them from objects with tilde that appeared in $\Delta S$ in eq. (\ref{eq:DeltaSlargeR}). They are different because, as we found in Sec. \ref{subsec:DeltaH}, $\Delta \langle H \rangle$ in general gives nonzero terms of this form in the large $R$ expansion.

For $d=2$, (\ref{t}) gives
\be\label{eq:Srel2}
(R S_\text{rel}'(R))' \ge 0\,.
\ee
Let us introduce the defect C-function
\be
\Delta C_\text{defect}(R) \equiv - R S_\text{rel}'(R)\,.
\ee
At large $R$, this gives the coefficient of the logarithmic term in the relative entropy. Eq. (\ref{eq:Srel2}) then says that $\Delta C_\text{defect}'(R)  \le 0$ along defect RG flows, for a defect with $d=2$ in a $D$-dimensional CFT. In terms of the parametrization in (\ref{eq:Srelprime}), the universal term decreases between the UV and IR fixed point, $  A'_{IR} \le A'_{UV}$ for $d=2$. This inequality was proved before in \cite{jensen2016constraint} using dilaton methods.

For $d=3$, we define the defect F-function
\be
\Delta F_\text{defect}(R) \equiv-R S'_\text{rel}(R)+S_\text{rel}(R)
\ee
and (\ref{t}) gives
\be
(\Delta F_\text{defect}(R))'  \le 0\,.
\ee
Therefore this quantity decreases monotonically and coincides with $\Delta F'$ at fixed points. This gives the F-theorem for defects,
 \be
F'_{UV}\ge F'_{IR}\,,
\ee
 a new result of our work which has not been proven before by other methods.

For $d=4$, replacing (\ref{eq:Srelprime}) into (\ref{t}), the leading area term cancels, and the logarithmic contribution gives $\Delta A' \le 0$, namely
\be
A'_{IR} \le A'_{UV}\;,\;\text{for}\;\;d=4\,.
\ee
We hence establish the entropic A-theorem for defects. It was proved using dilaton methods in \cite{wang2022defect}.

We conclude that the formula (\ref{t}) provides a unified proof for defect irreversibility theorems for $d=2,\, 3,\,4$. 

For $d \ge 5$, the second derivative inequality does not allow to access the universal terms; as in QFT without defects, it remains an open problem to establish inequalities for these terms. Instead, for general $d$ we obtain two inequalities for non-universal term. First, the interpolating quantity
\be
\Delta \mu'_{d-2}(R) \equiv - \frac{S_\text{rel}'(R)}{(d-2) R^{d-3}}
\ee
decreases due to (\ref{t}). This is an ``area theorem'' for defects; the analog theorem for QFTs without defects was established in \cite{Casini:2014yca, Casini:2016udt}. The second inequality comes from evaluating (\ref{t}) in the large $R$ limit (\ref{eq:Srelprime}), which gives
\be
\Delta \mu'_{d-4} \ge 0\,.
\ee
So the coefficient of $R^{d-4}$ in the relative entropy also changes with a fixed sign along RG flows. It would be interesting to understand if there is a gravitational interpretation for these results (as in \cite{Casini:2014yca}). It is also natural to conjecture that
\be
(-1)^k \Delta \mu'_{d-2k} \ge 0\,,
\ee
but, again, the SSA  or QNEC inequalities are not powerful enough to establish this.\footnote{Additional evidence for related inequalities comes from holographic theories \cite{Daguerre:2022uxt}.}

\section{Another look at defect irreversibility}\label{sec:another}

Sec. \ref{sec:defectirrev} presented a simple proof of the defect irreversibility formula by using infinitesimal deformations and the QNEC. In this section we will discuss an alternative proof that provides additional intuition for the necessity of the QNEC. This will also make contact with older proofs of irreversibility theorems using SSA. 

In order to establish the main points, it will be sufficient to restrict to the simplest case of a defect with $d=2$ and $D=3$. Let us consider the setup of boosted spheres used originally to prove the C-theorem (without defects) in \cite{Casini:2004bw}. We have two boosted spheres $X_1$ and $X_2$ of diameter $\sqrt{R r}$, centered on the defect and extending into the bulk, as in Fig. \ref{fig:unamasa} a). The SSA inequality for the entanglement entropy is
\be\label{eq:S2d}
S(X_1) + S(X_2) \ge S(X_1 \cup X_2) + S(X_1 \cap X_2)\,.
\ee
For $d=D=2$ (i.e. no defect), $X_1 \cup X_2$ defines the causal diamond for an interval of size $R$, and $X_1 \cap X_2$ corresponds to an interval of size $r$. Then this inequality reduces to $2 S(\sqrt{r R}) \ge S(R) + S(r)$, whose infinitesimal version $r \to R$ gives the C-theorem inequality $R S''(R)+S'(R) \le 0$. However, when $D> d$ the union and intersection of spheres no longer give causal diamonds of spheres. See figure \ref{fig:unamasa} b).  As it stands, then, (\ref{eq:S2d}) does not give the C-theorem for defects.\footnote{ This is why the attempts in \cite{Azeyanagi:2007qj, Yuan:2022oeo} do not work. In this last work in particular, the authors try to avoid defining regions in the bulk, and study a definition of entropy for regions on the defect. However, the defect theory is in general non local and SSA cannot be applied in the usual way.} 

This problem is solved in two steps. First, using the Markov property of the CFT vacuum, the SSA inequality is written in terms of the relative entropy $S_\text{rel}(\rho|\sigma)$,
\be\label{eq:S2d2}
2 S_\text{rel}(\sqrt{r R})  \le S_\text{rel}(X_1 \cup X_2) + S_\text{rel}(X_1 \cap X_2)\,.
\ee
The second step is to use the QNEC to prove that
\be\label{eq:Sreltemp2}
S_\text{rel}(X_1 \cup X_2) + S_\text{rel}(X_1 \cap X_2) \le S_\text{rel}(R) +S_\text{rel}(r)\,.
\ee
In Fig. \ref{fig:unamasa}, this is shown in going from b) to c). A simple way to see this geometrically is to conformally map the regions from the light-cone to the null plane. The boundaries of the regions of the null plane correspond to two parabolas and two lines, and  (\ref{eq:Sreltemp2}) can be obtained as the finite version of the QNEC convexity property (\ref{ff}). Plugging (\ref{eq:Sreltemp2}) into (\ref{eq:S2d2}) then gives
\be
2 S_\text{rel}(\sqrt{r R})  \le S_\text{rel}(R) +S_\text{rel}(r)\,,
\ee
whose infinitesimal version is the desired defect C-theorem.

The generalization to arbitrary $d$ and $D$ involves applying the QNEC to a large number $N$ of boosted spheres, as in \cite{Casini:2012ei, casini2017markov}. As in the simple case that we just discussed, we can only perform boosts that keep the defect invariant. The impossibility of performing boosts along all the $D$-dimensional spacetime implies that unions and intersections of spheres will not converge to spheres, and this is the stage at which the QNEC enters. 

In more detail, the SSA inequality reads
\bea
N S_\text{rel}(\sqrt{R r})\le S_\text{rel}(\cup_i X_i)+S_\text{rel}(\cup_{\{ ij\}}(X_i\cap X_j))+ \dots+ S_\text{rel}(\cap_i X_i)\,.
\label{mSSA2}
\eea
As discussed before, the problem comes from the right hand side of (\ref{mSSA2}): the unions and intersections of the $X_i$ correspond to causal developments with large wiggles. They do not converge to causal diamonds of spheres.

At large $N$, these regions $\cup_i X_i, \,\cup_{\{ ij\}}(X_i\cap X_j)\,\ldots$, intersect with the defect, defining (wiggly) spheres on approximately constant $t$ planes, with radius $l$ between $r$ and $R$. The density of these spheres is \cite{casini2017markov}
\be
\beta(l)= \frac{2^{d-3}\Gamma[(d-1)/2](rR)^{\frac{d-2}2}((l-r)(R-l))^{\frac{d-4}2}}{\sqrt{\pi} \Gamma[(d-2)/2]\,l^{\,d-2}(R-l)^{d-3}}\,.
\ee
Subtracting their contribution from both sides of (\ref{mSSA2}) gives
\bea
 S_\text{rel}(\sqrt{R r})-\int_r^R dl \beta(l) S_\text{rel}(l)&\le& \frac{1}{N}\left[S_\text{rel}(\cup_i X_i)+S_\text{rel}(\cup_{\{ ij\}}(X_i\cap X_j))+ \dots+ S_\text{rel}(\cap_i X_i)\right]\nonumber\\
&&\qquad-\int_r^R dl \beta(l) S_\text{rel}(l)\,.
\eea
By the QNEC, the right hand side turns out to be negative,
\be
\frac{1}{N}\left[S_\text{rel}(\cup_i X_i)+S_\text{rel}(\cup_{\{ ij\}}(X_i\cap X_j))+ \dots+ S_\text{rel}(\cap_i X_i)\right]-\int_r^R dl \beta(l) S_\text{rel}(l) \le 0\,,
\ee
and the infinitesimal version of the resulting inequality,
\be
S_\text{rel}(\sqrt{R r})-\int_r^R dl \beta(l) S_\text{rel}(l) \le 0
\ee
gives (\ref{t}).

Summarizing, the inequality follows from applying SSA in the positions of the spheres on the defect, which is always a valid operation, plus the QNEC for the bulk on the null cone, where it is valid due to conformal invariance of the theory.

\section{Holographic connection of irreversibility with and without defects}\label{sec:holography}

The irreversibility theorems for theories without defect must also hold for holographic theories. They can be proved directly from the behaviour of the bulk metric comparing the end points of a holographic RG flow \cite{myers2011holographic, Daguerre:2022uxt}. However, as these theorems in their entropic version arise from the SSA property of the entropy, a holographic proof also follows  from a holographic proof of SSA for holographic entanglement entropy. This proof follows from the maximin construction for the extremal surfaces of minimal area, and uses the bulk null energy condition \cite{wall2014maximin}.

For most holographic RGs this suffices. However, we may wonder what happens for a holographic RG flow where Newton's constant is kept invariant. This type of flow can appear for instance from double trace deformations \cite{Aharony:2001pa, Witten:2001ua}. In this case, the bulk remains pure AdS (to lowest order in $N$), but there is a bulk scalar field whose boundary condition changes between Newman and Dirichlet. From the bulk point of view, this looks precisely like a defect flow. 

It is then interesting to follow  entropic proof of irreversibility for the boundary QFT (without defects) to understand the connection with the bulk interpretation as defect flow irreversibility. An important point is that the changes in the entropy we want to follow are order $N^0$ and correspond to the quantum part in the entropy. This is equal to the bulk quantum entanglement entropy across the minimal surface \cite{faulkner2013quantum}. In fact, the larger $\sim G^{-1}$ term in the entropy, being proportional to the area in AdS, gives the entropy of a CFT, and exactly cancels in the construction of the irreversibility theorems.  Therefore, the relevant part of the entropy  consists only on the quantum part. The construction of the boundary irreversibility theorem then exactly matches in the bulk our construction in this paper to prove the irreversibility of the defect flow, specialising to the case of codimension one. The validity of the boundary irreversibility theorem depends then on the validity of the defect irreversibility in the bulk. As such, the boundary proof of irreversibility, that essentially depends on SSA, must use the QNEC in the bulk. 

More generally, one may wonder why the holographic entanglement entropy still obeys the SSA property when the quantum correction is taken into account. Quantum effects may produce violations of the NEC for example, and the classical proof must be generalized. This generalization was accomplished in \cite{akers2020quantum}, where the key ingredient was to replace the NEC by the quantum focusing conjecture \cite{bousso2016quantum}. In the limit of the double trace deformations described above, and where the geometry is fixed, this quantum focusing conjecture turns into the QNEC. This closes the circle of ideas around the connection between the two irreversibility theorems in holographic theories. The proof of holographic SSA (including the quantum correction) in \cite{akers2020quantum} necessitates of the quantum focusing conjecture to be able to rearrange the position of the  surfaces in the bulk such as to match intersections and unions with the minimal surfaces corresponding to the boundary intersections and unions. This exactly parallels our necessity of rearranging surfaces to modify the form of the SSA such that the inequality involves only regions (spheres) that depend exclusively on their boundary support. Furthermore, for codimension one cases, the modular Hamiltonian contribution vanishes, and hence $S_\text{rel}(R) = -\Delta S(R)$.        

 We notice that in the above argument we have not used that the bulk theory is a CFT. If it is a CFT the AdS metric can be conformally transformed to Minkowski and indeed the bulk interpretation is equivalent to a defect flow in a CFT. However, we have in more generality the irreversibility of the universal terms in the relative entropy for generic QFT in AdS with generic boundary conditions imposed on the AdS boundary.\footnote{See \cite{benedetti2023mutual} for an explicit calculation of the universal terms for free massive fields in AdS with Neumann and Dirichlet boundary conditions.} In that case we have to use the usual bulk QNEC for null surfaces without expansion, corresponding to the causal wedge of null cones in the boundary. This constitutes a different form of defect irreversibility from the previous one considered in the literature.

\section*{Acknowledgments} 

We thank discussions with Eduardo Test\'e around the calculations presented in section \ref{sec:irrev}. HC and GT are supported by
CONICET (PIP grant 11220200101008CO), ANPCyT (PICT 2018-2517), CNEA, and Instituto Balseiro, Universidad Nacional de Cuyo. HC acknowledges an ``It From Qubit" grant of the Simons Foundation. ISL would like to thank IB and ICTP for hospitality.

\bibliography{EE}{}

\providecommand{\href}[2]{#2}\begingroup\raggedright\begin{thebibliography}{10}

\bibitem{affleck1991universal}
I.~Affleck and A.~W. Ludwig, ``Universal noninteger ‘‘ground-state
  degeneracy’’in critical quantum systems,'' {\em Physical Review Letters}
  {\bfseries 67} no.~2, (1991) 161.

\bibitem{friedan2004boundary}
D.~Friedan and A.~Konechny, ``Boundary entropy of one-dimensional quantum
  systems at low temperature,'' {\em Physical review letters} {\bfseries 93}
  no.~3, (2004) 030402.

\bibitem{Zamolodchikov:1986gt}
A.~B. Zamolodchikov, ``{Irreversibility of the Flux of the Renormalization
  Group in a 2D Field Theory},'' {\em JETP Lett.} {\bfseries 43} (1986)
  730--732.
[Pisma Zh. Eksp. Teor. Fiz.43,565(1986)].

\bibitem{Cardy:1988cwa}
J.~L. Cardy, ``{Is There a c Theorem in Four-Dimensions?},''
\href{http://dx.doi.org/10.1016/0370-2693(88)90054-8}{{\em Phys. Lett.}
  {\bfseries B215} (1988) 749--752}.

\bibitem{holzhey1994geometric}
C.~Holzhey, F.~Larsen, and F.~Wilczek, ``Geometric and renormalized entropy in
  conformal field theory,'' {\em Nuclear physics b} {\bfseries 424} no.~3,
  (1994) 443--467.

\bibitem{solodukhin2008entanglement}
S.~N. Solodukhin, ``Entanglement entropy, conformal invariance and extrinsic
  geometry,'' {\em Physics Letters B} {\bfseries 665} no.~4, (2008) 305--309.

\bibitem{casini2011towards}
H.~Casini, M.~Huerta, and R.~C. Myers, ``Towards a derivation of holographic
  entanglement entropy,'' {\em Journal of High Energy Physics} {\bfseries 2011}
  no.~5, (2011) 1--41.

\bibitem{Myers:2010xs}
R.~C. Myers and A.~Sinha, ``{Seeing a c-theorem with holography},''
  \href{http://dx.doi.org/10.1103/PhysRevD.82.046006}{{\em Phys. Rev.}
  {\bfseries D82} (2010) 046006},
\href{http://arxiv.org/abs/1006.1263}{{\ttfamily arXiv:1006.1263 [hep-th]}}.

\bibitem{Jafferis:2011zi}
D.~L. Jafferis, I.~R. Klebanov, S.~S. Pufu, and B.~R. Safdi, ``{Towards the
  F-Theorem: N=2 Field Theories on the Three-Sphere},''
  \href{http://dx.doi.org/10.1007/JHEP06(2011)102}{{\em JHEP} {\bfseries 06}
  (2011) 102},
\href{http://arxiv.org/abs/1103.1181}{{\ttfamily arXiv:1103.1181 [hep-th]}}.

\bibitem{calabrese2004entanglement}
P.~Calabrese and J.~Cardy, ``Entanglement entropy and quantum field theory,''
  {\em Journal of Statistical Mechanics: Theory and Experiment} {\bfseries
  2004} no.~06, (2004) P06002.

\bibitem{Jensen:2018rxu}
K.~Jensen, A.~O'Bannon, B.~Robinson, and R.~Rodgers, ``{From the Weyl Anomaly
  to Entropy of Two-Dimensional Boundaries and Defects},''
  \href{http://dx.doi.org/10.1103/PhysRevLett.122.241602}{{\em Phys. Rev.
  Lett.} {\bfseries 122} no.~24, (2019) 241602},
  \href{http://arxiv.org/abs/1812.08745}{{\ttfamily arXiv:1812.08745
  [hep-th]}}.

\bibitem{kobayashi2019towards}
N.~Kobayashi, T.~Nishioka, Y.~Sato, and K.~Watanabe, ``Towards a c-theorem in
  defect cft,'' {\em Journal of High Energy Physics} {\bfseries 2019} no.~1,
  (2019) 1--47.

\bibitem{komargodski2011renormalization}
Z.~Komargodski and A.~Schwimmer, ``On renormalization group flows in four
  dimensions,'' {\em Journal of High Energy Physics} {\bfseries 2011} no.~12,
  (2011) 1--20.

\bibitem{jensen2016constraint}
K.~Jensen and A.~O’Bannon, ``Constraint on defect and boundary
  renormalization group flows,'' {\em Physical Review Letters} {\bfseries 116}
  no.~9, (2016) 091601.

\bibitem{Shachar:2022fqk}
T.~Shachar, R.~Sinha, and M.~Smolkin, ``{RG flows on two-dimensional spherical
  defects},'' \href{http://arxiv.org/abs/2212.08081}{{\ttfamily
  arXiv:2212.08081 [hep-th]}}.

\bibitem{wang2022defect}
Y.~Wang, ``Defect a-theorem and a-maximization,'' {\em Journal of High Energy
  Physics} {\bfseries 2022} no.~2, (2022) 1--46.

\bibitem{cuomo2022renormalization}
G.~Cuomo, Z.~Komargodski, and A.~Raviv-Moshe, ``Renormalization group flows on
  line defects,'' {\em Physical Review Letters} {\bfseries 128} no.~2, (2022)
  021603.

\bibitem{Casini:2004bw}
H.~Casini and M.~Huerta, ``{A Finite entanglement entropy and the c-theorem},''
  \href{http://dx.doi.org/10.1016/j.physletb.2004.08.072}{{\em Phys. Lett.}
  {\bfseries B600} (2004) 142--150},
\href{http://arxiv.org/abs/hep-th/0405111}{{\ttfamily arXiv:hep-th/0405111
  [hep-th]}}.

\bibitem{Casini:2012ei}
H.~Casini and M.~Huerta, ``{On the RG running of the entanglement entropy of a
  circle},'' \href{http://dx.doi.org/10.1103/PhysRevD.85.125016}{{\em Phys.
  Rev.} {\bfseries D85} (2012) 125016},
\href{http://arxiv.org/abs/1202.5650}{{\ttfamily arXiv:1202.5650 [hep-th]}}.

\bibitem{casini2017markov}
H.~Casini, E.~Test{\'e}, and G.~Torroba, ``Markov property of the conformal
  field theory vacuum and the a theorem,'' {\em Physical review letters}
  {\bfseries 118} no.~26, (2017) 261602.

\bibitem{casini2016g}
H.~Casini, I.~S. Landea, and G.~Torroba, ``The g-theorem and quantum
  information theory,'' {\em Journal of High Energy Physics} {\bfseries 2016}
  no.~10, (2016) 1--34.

\bibitem{casini2023entropic}
H.~Casini, I.~S. Landea, and G.~Torroba, ``Entropic g theorem in general
  spacetime dimensions,'' {\em Physical Review Letters} {\bfseries 130} no.~11,
  (2023) 111603.

\bibitem{casini2017relative}
H.~Casini, E.~Test{\'e}, and G.~Torroba, ``Relative entropy and the rg flow,''
  {\em Journal of High Energy Physics} {\bfseries 2017} no.~3, (2017) 1--24.

\bibitem{casini2019irreversibility}
H.~Casini, I.~S. Landea, and G.~Torroba, ``Irreversibility in quantum field
  theories with boundaries,'' {\em Journal of High Energy Physics} {\bfseries
  2019} no.~4, (2019) 1--18.

\bibitem{bousso2016quantum}
R.~Bousso, Z.~Fisher, S.~Leichenauer, and A.~C. Wall, ``Quantum focusing
  conjecture,'' {\em Physical Review D} {\bfseries 93} no.~6, (2016) 064044.

\bibitem{koeller2016holographic}
J.~Koeller and S.~Leichenauer, ``Holographic proof of the quantum null energy
  condition,'' {\em Physical Review D} {\bfseries 94} no.~2, (2016) 024026.

\bibitem{casini2017modular}
H.~Casini, E.~Test{\'e}, and G.~Torroba, ``Modular hamiltonians on the null
  plane and the markov property of the vacuum state,'' {\em Journal of Physics
  A: Mathematical and Theoretical} {\bfseries 50} no.~36, (2017) 364001.

\bibitem{casini2018all}
H.~Casini, E.~Test{\'e}, and G.~Torroba, ``All the entropies on the
  light-cone,'' {\em Journal of High Energy Physics} {\bfseries 2018} no.~5,
  (2018) 5.

\bibitem{mezei2015entanglement}
M.~Mezei, ``Entanglement entropy across a deformed sphere,'' {\em Physical
  Review D} {\bfseries 91} no.~4, (2015) 045038.

\bibitem{faulkner2016shape}
T.~Faulkner, R.~G. Leigh, and O.~Parrikar, ``Shape dependence of entanglement
  entropy in conformal field theories,'' {\em Journal of High Energy Physics}
  {\bfseries 2016} no.~4, (2016) 1--39.

\bibitem{petz2007quantum}
D.~Petz, {\em Quantum information theory and quantum statistics}.
\newblock Springer Science \& Business Media, 2007.

\bibitem{bousso2016proof}
R.~Bousso, Z.~Fisher, J.~Koeller, S.~Leichenauer, and A.~C. Wall, ``Proof of
  the quantum null energy condition,'' {\em Physical Review D} {\bfseries 93}
  no.~2, (2016) 024017.

\bibitem{balakrishnan2019general}
S.~Balakrishnan, T.~Faulkner, Z.~U. Khandker, and H.~Wang, ``A general proof of
  the quantum null energy condition,'' {\em Journal of High Energy Physics}
  {\bfseries 2019} no.~9, (2019) 1--86.

\bibitem{balakrishnan2022entropy}
S.~Balakrishnan, V.~Chandrasekaran, T.~Faulkner, A.~Levine, and
  A.~Shahbazi-Moghaddam, ``Entropy variations and light ray operators from
  replica defects,'' {\em Journal of High Energy Physics} {\bfseries 2022}
  no.~9, (2022) 1--42.

\bibitem{leichenauer2018energy}
S.~Leichenauer, A.~Levine, and A.~Shahbazi-Moghaddam, ``Energy density from
  second shape variations of the von neumann entropy,'' {\em Physical Review D}
  {\bfseries 98} no.~8, (2018) 086013.

\bibitem{wall2012proof}
A.~C. Wall, ``Proof of the generalized second law for rapidly changing fields
  and arbitrary horizon slices,'' {\em Physical Review D} {\bfseries 85}
  no.~10, (2012) 104049.

\bibitem{Longo:2018obd}
R.~Longo, ``{Entropy distribution of localised states},''
  \href{http://dx.doi.org/10.1007/s00220-019-03332-8}{{\em Commun. Math. Phys.}
  {\bfseries 373} no.~2, (2019) 473--505},
  \href{http://arxiv.org/abs/1809.03358}{{\ttfamily arXiv:1809.03358
  [hep-th]}}.

\bibitem{Casini:2016fgb}
H.~Casini, I.~S. Landea, and G.~Torroba, ``{The g-theorem and quantum
  information theory},'' \href{http://dx.doi.org/10.1007/JHEP10(2016)140}{{\em
  JHEP} {\bfseries 10} (2016) 140},
\href{http://arxiv.org/abs/1607.00390}{{\ttfamily arXiv:1607.00390 [hep-th]}}.

\bibitem{billo2016defects}
M.~Billo, V.~Gon{\c{c}}alves, E.~Lauria, and M.~Meineri, ``Defects in conformal
  field theory,'' {\em Journal of High Energy Physics} {\bfseries 2016} no.~4,
  (2016) 1--56.

\bibitem{mezei2019quantum}
M.~Mezei and J.~Virrueta, ``The quantum null energy condition and entanglement
  entropy in quenches,'' {\em arXiv preprint arXiv:1909.00919} (2019) .

\bibitem{Casini:2014yca}
H.~Casini, F.~D. Mazzitelli, and E.~Teste, ``{Area terms in entanglement
  entropy},'' \href{http://dx.doi.org/10.1103/PhysRevD.91.104035}{{\em Phys.
  Rev.} {\bfseries D91} no.~10, (2015) 104035},
\href{http://arxiv.org/abs/1412.6522}{{\ttfamily arXiv:1412.6522 [hep-th]}}.

\bibitem{Casini:2016udt}
H.~Casini, E.~Teste, and G.~Torroba, ``{Relative entropy and the RG flow},''
  \href{http://dx.doi.org/10.1007/JHEP03(2017)089}{{\em JHEP} {\bfseries 03}
  (2017) 089},
\href{http://arxiv.org/abs/1611.00016}{{\ttfamily arXiv:1611.00016 [hep-th]}}.

\bibitem{Daguerre:2022uxt}
L.~Daguerre, M.~Ginzburg, and G.~Torroba, ``{Holographic entanglement entropy
  inequalities beyond strong subadditivity},''
  \href{http://dx.doi.org/10.1007/JHEP10(2022)199}{{\em JHEP} {\bfseries 10}
  (2022) 199}, \href{http://arxiv.org/abs/2208.03334}{{\ttfamily
  arXiv:2208.03334 [hep-th]}}.

\bibitem{Azeyanagi:2007qj}
T.~Azeyanagi, A.~Karch, T.~Takayanagi, and E.~G. Thompson, ``{Holographic
  calculation of boundary entropy},''
  \href{http://dx.doi.org/10.1088/1126-6708/2008/03/054}{{\em JHEP} {\bfseries
  03} (2008) 054}, \href{http://arxiv.org/abs/0712.1850}{{\ttfamily
  arXiv:0712.1850 [hep-th]}}.

\bibitem{Yuan:2022oeo}
M.-K. Yuan and Y.~Zhou, ``{Defect Localized Entropy: Renormalization Group and
  Holography},'' \href{http://arxiv.org/abs/2209.08835}{{\ttfamily
  arXiv:2209.08835 [hep-th]}}.

\bibitem{myers2011holographic}
R.~C. Myers and A.~Sinha, ``Holographic c-theorems in arbitrary dimensions,''
  {\em Journal of High Energy Physics} {\bfseries 2011} no.~1, (2011) 1--53.

\bibitem{wall2014maximin}
A.~C. Wall, ``Maximin surfaces, and the strong subadditivity of the covariant
  holographic entanglement entropy,'' {\em Classical and Quantum Gravity}
  {\bfseries 31} no.~22, (2014) 225007.

\bibitem{Aharony:2001pa}
O.~Aharony, M.~Berkooz, and E.~Silverstein, ``{Multiple trace operators and
  nonlocal string theories},''
  \href{http://dx.doi.org/10.1088/1126-6708/2001/08/006}{{\em JHEP} {\bfseries
  08} (2001) 006}, \href{http://arxiv.org/abs/hep-th/0105309}{{\ttfamily
  arXiv:hep-th/0105309}}.

\bibitem{Witten:2001ua}
E.~Witten, ``{Multitrace operators, boundary conditions, and AdS / CFT
  correspondence},'' \href{http://arxiv.org/abs/hep-th/0112258}{{\ttfamily
  arXiv:hep-th/0112258}}.

\bibitem{faulkner2013quantum}
T.~Faulkner, A.~Lewkowycz, and J.~Maldacena, ``Quantum corrections to
  holographic entanglement entropy,'' {\em Journal of High Energy Physics}
  {\bfseries 2013} no.~11, (2013) 1--18.

\bibitem{akers2020quantum}
C.~Akers, N.~Engelhardt, G.~Penington, and M.~Usatyuk, ``Quantum maximin
  surfaces,'' {\em Journal of High Energy Physics} {\bfseries 2020} no.~8,
  (2020) 1--43.

\bibitem{benedetti2023mutual}
V.~Benedetti, H.~Casini, and P.~J. Martinez, ``Mutual information of
  generalized free fields,'' {\em Physical Review D} {\bfseries 107} no.~4,
  (2023) 046003.

\end{thebibliography}\endgroup
\bibliographystyle{utphys}

\end{document}